\documentclass[aps,prl,twocolumn,superscriptaddress]{revtex4}

\usepackage{graphicx}

\begin{document}

{\bf Comment on ``Quantum and Classical Glass Transitions in LiHo$_{x}$Y$_{1-x}$F$_4$''}

We show in this comment that the claim by Ancona-Torres et al. (A.-T. et al.) of equilibrium quantum and classical phase transitions in the LiHo$_x$Y$_{1-x}$F$_4$ $x$=0.167 and $x$=0.198 systems\cite{rosenbaum} are not supported by reliable analyses.

Firstly, the divergence of the non-linear susceptibility $\chi_{nl}$ or its lowest order term $\chi_3$, which would evidence the classical phase transition, is only claimed by A.-T. et al., but not scientifically evidenced by e.g. rigorous scaling analysis. Such analyses have successfully been employed in a number of studies, e.g. \cite{bb,roland-jpsj} in order to evidence the existence of a spin glass phase transition, as well as in our recent study to evidence the absence of equilibrium phase transitions in LiHo$_x$Y$_{1-x}$F$_4$ ($x$=0.167) \cite{us}.

\begin{figure}[hb]
\includegraphics[width=0.46\textwidth]{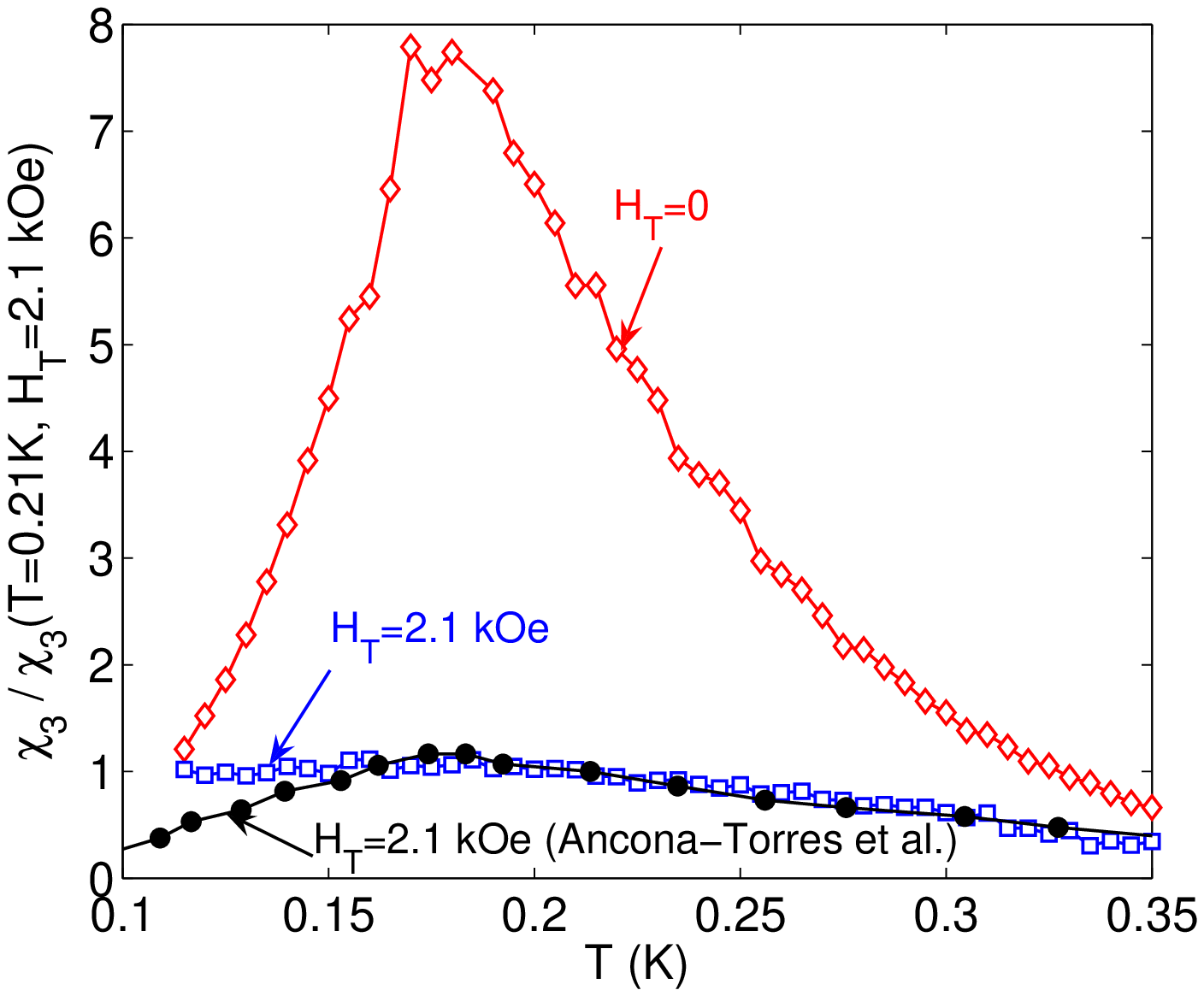}
\caption{Temperature dependence of $\chi_3$. We have plotted our data from Ref.~\onlinecite{us}, recorded in zero and 2.1 kOe transverse magnetic fields $H_T$ (open markers), as well as that presented by A.-T. et al.\cite{rosenbaum} for $H_T$=2.1 kOe (filled markers; A.-T. et al. do not show $H_T$=0 data). In order to compare the curves recorded in $H_T$=2.1 kOe more clearly, we have normalized $\chi_3$ by its respective magnitude at $T$= 0.21 K and $H_T$=2.1 kOe.}
\label{fig}
\end{figure}

In Fig.~\ref{fig}, we compare our $\chi_3(T)$ to that of A.-T. et al.'s recorded with low longitudinal fields. We find that our $\chi_3(T)$ data recorded in a transverse field $H_T$ of 2.1 kOe is relatively similar to that of A.-T. et al., in the same field (please note that for these static analysis, only the data above the cusp in the susceptibility need to be considered). This casts some shadow on the correctness of the conclusions of A.-T. et al. about the divergence of the non-linear susceptibility. Our $\chi_3(T)$  data, recorded in zero transverse field, also reproduced in the figure, ``looks more divergent'' that the one in $H_T=2.1$~kOe. Yet we have shown that this $\chi_3(T,H_T=0)$ data does not diverge, using rigorous scaling laws\cite{us}. It thus seems unlikely that A.-T. et al.'s $\chi_3(T,H_T=2.1$ kOe) diverges as they claim. Although it is evident, we have checked that our $\chi_3(T,H_T=2.1$) kOe does not diverge either.

Next, the spin-glass transition temperature is defined from dynamical susceptibility measurements by A.-T. et al. using hand-waving arguments as the temperature at which $\alpha \to 0$, where $\chi'' \sim \omega^\alpha$, e.g. as the highest temperature at which $\chi''(\omega)$ is flat if $\omega \to 0$ (within the limited experimental time window).

A rigorous way to evidence a spin-glass phase transition is to perform a so-called dynamical scaling analysis\cite{roland-jpsj,bbnew} in which the divergence of the characteristic relaxation time $\tau_c=\tau_0(T/T_g-1)^{-z\nu}$ is confirmed. 
The characteristic relaxation time can be defined as $\tau_c = \lim_{\omega \to 0} \frac{\chi''}{\omega \chi'}$ \cite{ogielski85} within a limited temperature range above $T_g$. However the analysis of the critical slowing down is non-trivial in the present system, as the fastest timescale of the relaxation $\tau_0$ is temperature dependent and much slower  ($\sim 1$~ms at $T=0.13$~K as can be seen in Fig. 5(d) of \cite{rosenbaum}) than the atomic flip time ($\sim 10^{-13}$~s) in canonical spin glasses.

\acknowledgments

P.E.J. and R.M. acknowledge the Swedish Research Council (VR) for financial support. B.B and A.M.T. acknowledge the European contract INTAS-2003/05-51-4943.\\

\noindent
P. E. J{\"o}nsson,$^1$ R. Mathieu,$^2$ W. Wernsdorfer,$^3$ A. M. Tkachuk,$^4$ and B. Barbara$^3$\\
{\small
$^1$ Department of Physics, Uppsala University, Box 530, SE-751 21 Uppsala, Sweden\\
$^2$ Department of Engineering Sciences, Uppsala University, Box 534, SE-751 21 Uppsala, Sweden\\
$^3$ Institut N\'eel, CNRS/UJF, 25 avenue des Martyrs, BP166, 38042 Grenoble Cedex 9, France\\
$^4$All-Russia S.I. Vavilov State Optical Institute, St. Petersburg 199034, Russia}

\end{document}